\documentclass[pra, twocolumn, nofootinbib]{revtex4-1}

\usepackage{custom}
\usepackage[hidelinks]{hyperref}
\def\name{Alexander R. H. Smith}
\hypersetup{
  colorlinks = False,
  urlcolor = blue,
  linkcolor = black, 
  pdfauthor = {\name},
  pdftitle = {\name: Quantum time dilation},
}

\begin{document}

 \title{{\Large Quantum time dilation: A new test of relativistic quantum theory}  \\ \vspace{2pt}
 { \normalsize  \it{Essay written for the Gravity Research Foundation 2020 Awards for Essays on Gravitation} }}
 
 \author{Alexander R. H. Smith}
\email[]{alexander.r.smith@dartmouth.edu}
\affiliation{Department of Physics and Astronomy, Dartmouth College, Hanover, New Hampshire 03755, USA}

%
\date{\today}

\begin{abstract}
A novel quantum time dilation effect is shown to arise when a clock moves in a quantum superposition of two relativistic velocities. This effect is argued to be measurable using existing atomic interferometry techniques, potentially offering a new test of relativistic quantum theory.
\end{abstract}
\maketitle


John Wheeler advocated a radically conservative approach to physics: Insist on adhering to well-established physical laws (be conservative), but follow those laws into their most extreme domains (be radical), where unexpected insights into nature might be found~\cite{misnerJohnWheelerRelativity2009}. Our most well-established  laws are encapsulated in quantum mechanics and general relativity, which respectively predict the phenomena of superposition and special relativistic time dilation. Taking Wheeler's advice and following these theories to their extremes begs the questions:
\begin{quote}
\emph{What  time dilation is observed by a clock that moves in a quantum superposition of two different relativistic velocities? And supposing a novel quantum time dilation effect does exist, is it experimentally observable?}\footnote{This essay is based on work that appears in Ref.~\cite{smithRelativisticQuantumClocks2019}. Using internal degrees of freedom as quantum clocks to examine relativistic effects was introduced in Ref.~\cite{zychQuantumInterferometricVisibility2011a} and related effects have been examined in Refs.~\cite{vedralSchrodingerCatMeets2008,ruizEntanglementQuantumClocks2017,
bushevSingleElectronRelativistic2016,zychGravitationalMassComposite2019a,khandelwalGeneralRelativisticTime2019,paigeClassicalNonclassicalTime2020,hohnSwitchingInternalTimes2019a,hoehnHowSwitchRelational2018,lorianiInterferenceClocksQuantum2019a,TimeReferenceFrames2019}.}
\end{quote}

To answer these questions, consider a relativistic particle of mass $m$, with an internal degree of freedom that functions as a clock $C$. The action for such a particle is
\begin{align}
\mathcal{S} =  \int d \tau \,  \left( -m c^2 + L_C \right),\nn 
\end{align}
where $d\tau$ a differential amount of the particle's proper time and $L_C$ is the clock Lagrangian. This action is Lorentz invariant, which leads to a Hamiltonian constraint
\begin{align}
C_H \ce \eta^{\mu \nu} p_\mu p_\nu +  M^2 c^4 \approx 0,
\nn
\end{align}
where $\eta_{\mu \nu}$ is the Minkowski metric, $p_{\mu}$ are components of the particle's momentum four-vector with respect to an inertial frame, and $\approx$ denotes a weak equality which is  satisfied only when the equations of motion are satisfied. The mass function  $M \ce m + H_{C}/c^2$ has also been introduced, which is the sum of the rest mass $m$ and the mass $H_C/c^2$ associated with the internal clock energy, with $H_C$ being the Hamiltonian obtained via a Legendre transformation of $L_C$. This constraint can be factorized as $C_H = C_+ C_-$, where
\begin{align}
C_{\pm} \ce p_0 \pm \sqrt{ \boldsymbol{p}^2c^2 + M^2 c^4},\nn
\end{align}
and $p_0$ and $\boldsymbol{p}$ are respectively the energy and spatial components of the particle's four-momentum.

To quantize the theory, the Hamiltonian constraint is promoted to a constraint operator $\hat{C}_H$, which defines the physical states $\ket{\Psi} \in \mathcal{H}_{\rm phys}$ of the theory as those which satisfy \begin{align}
\hat{C}_H \ket{\Psi}  = \hat{C}_+ \hat{C}_- \ket{\Psi}  = 0. \nn
\end{align}
As a consequence of the factorization $\hat{C}_H = \hat{C}_+ \hat{C}_-$ and commutation relation $\big[\hat{C}_+ , \hat{C}_- \big]= 0$, the physical Hilbert space $\mathcal{H}_{\rm phys}$ decomposes into positive and negative frequency charge sectors, $\mathcal{H}_{\rm phys} = \mathcal{H}_{+} \oplus \mathcal{H}_{-}$, where $\mathcal{H}_{\pm}$ is defined as the subspace on which $C_{\pm} \ket{\Psi}  =0$~\mbox{\cite{hohnSwitchingInternalTimes2019a,hoehnHowSwitchRelational2018,hohnRelationalQuantumDynamics2020}}. In what follows  we restrict to the positive frequency sector and the constraint $\hat{C}_+ \ket{\Psi} =0$. 

This quantization scheme allows for the introduction of a proper time observable $T_C$ into the quantum theory.  Given that the clock Hamiltonian $\hat{H}_C$ generates an evolution of $C$ with respect to the proper time $\tau$ of the particle, $
\ket{\psi_{C}(\tau)} = e^{-i  \tau \hat{H}_C / \hbar}\ket{\psi_{C}(\tau_0)}$,  a proper time observable $T_C$ is defined as the observable whose expectation value on $\ket{\psi_{C}(\tau)}$ gives the best estimate of the parameter $\tau$. Such a time observable is not necessarily associated with a self-adjoint operator~\cite{buschTimeObservablesQuantum1994}, but is in general described by a positive operator valued measure (POVM). The proper time observable is defined by the set of rank-1 effect operators 
\begin{align}
T_C = \{ \ket{\tau}\!\bra{\tau}, \  \forall\, \tau \in G \}, \nn
\end{align}
where the (possibly generalized) clock states $\ket{\tau}$ yield the probability that $C$ indicates the time $\tau$ via the Born rule $\abs{\braket{\tau | \psi_C(\tau)}}^2$, and  $G$ is the set of proper times $\tau$ that can be indicated by $C$ upon a measurement of $T_C$. Furthermore, to yield the best estimate\footnote{Such a covariant proper time observable is an unbiased estimator that maximizes the Fisher information over all estimates of the parameter $\tau$ unitarily encoded in \mbox{$\ket{\psi_{C}(\tau)}$~\cite{holevoProbabilisticStatisticalAspects1982,braunsteinGeneralizedUncertaintyRelations1996}}. Furthermore, any estimate of $\tau$ must obey the Helstrom-Holevo lower bound~\mbox{\cite{helstromQuantumDetectionEstimation1976,holevoProbabilisticStatisticalAspects1982}}, which leads to a a generalized uncertainty relation between proper time and mass, $\Delta T_C \Delta \hat{M}  \geq \tfrac{\hbar}{2c^2  }$, where $\hat{M} \ce m + \hat{H}_{C}/c^2$ is a mass operator resulting from the quantization of the mass function $M$~\cite{smithRelativisticQuantumClocks2019}.} of $\tau$, the clock states must transform covariantly with respect to $\hat{H}_C$, meaning that $\ket{\tau+\tau'} = e^{-i\tau' \hat{H}_C /\hbar} \ket{\tau}$.

Now consider two  relativistic particles, $A$ and $B$, restricted to the positive sector of their respective physical Hilbert spaces, $\mathcal{H}^A_+$ and  $\mathcal{H}^B_+$, so that their joint Hilbert space is $\mathcal{H}_{\rm phys} = \mathcal{H}^A_+ \otimes \mathcal{H}^B_+$. The probability that clock $A$ reads the proper time $\tau_A$ when clock $B$ reads the proper time $\tau_B$ is given by the Born rule\footnote{The right hand side is to be evaluated in the kinematical inner product and the physical states normalized in an appropriate physical inner product~\cite{smithQuantizingTimeInteracting2019,hoehnTrinityRelationalQuantum2019}, in accordance with the Page-Wootters formalism~\cite{pageEvolutionEvolutionDynamics1983,woottersTimeReplacedQuantum1984}. One might worry that this probability is not  gauge invariant because it arises from the expectation values of kinematical observables. However, it has recently been shown that the Page-Wootters conditional probability distribution can be derived from relational Dirac observables and the standard Born rule, demonstrating that indeed this probability distribution is gauge independent~\cite{hoehnTrinityRelationalQuantum2019}.}
\begin{align}
\prob\!\left[ \tau_A  \mbox{ when }  \tau_B \right] =\frac{\braket{\Psi |\big(  \ket{\tau_A}\!\bra{\tau_A}  \otimes  \ket{\tau_B}\!\bra{\tau_B}  \big) | \Psi}}{\braket{\Psi|\big(   \ket{\tau_B}\!\bra{\tau_B} \big) |  \Psi}}. \label{probability}
\end{align}
This conditional probability depends sensitively on the quantum state of the center-of-mass of particles $A$ and $B$ as encoded in the physical state~$\ket{\Psi}$. Suppose that with respect to an inertial observer at Minkowski time $t=0$ the states of $A$ and $B$ are $\ket{\psi_A}$ and $\ket{\psi_B}$. Then the associated physical state is
\begin{align}
\ket{\Psi} = \int dt \, \ket{t} e^{-i (\hat{H}_A + \hat{H}_B) t/\hbar }\ket{\psi_A} \ket{\psi_B}, \nn
\end{align}
where $\hat{H}_i \ce \sqrt{ \hat{\boldsymbol{p}}_i^2c^2 + \hat{M}_i^2 c^4}$ for $i \in \{A,B\}$ is the Hamiltonian governing the evolution of both the internal clock degrees of freedom and external spatial/momentum degrees of freedom of the particles, and $\ket{\psi_i} = \ket{\psi_i^{\rm ext}}\ket{\psi_{C_i}}$ describes the joint state of these degrees of freedom.

\begin{figure}
\includegraphics[width= .45 \textwidth]{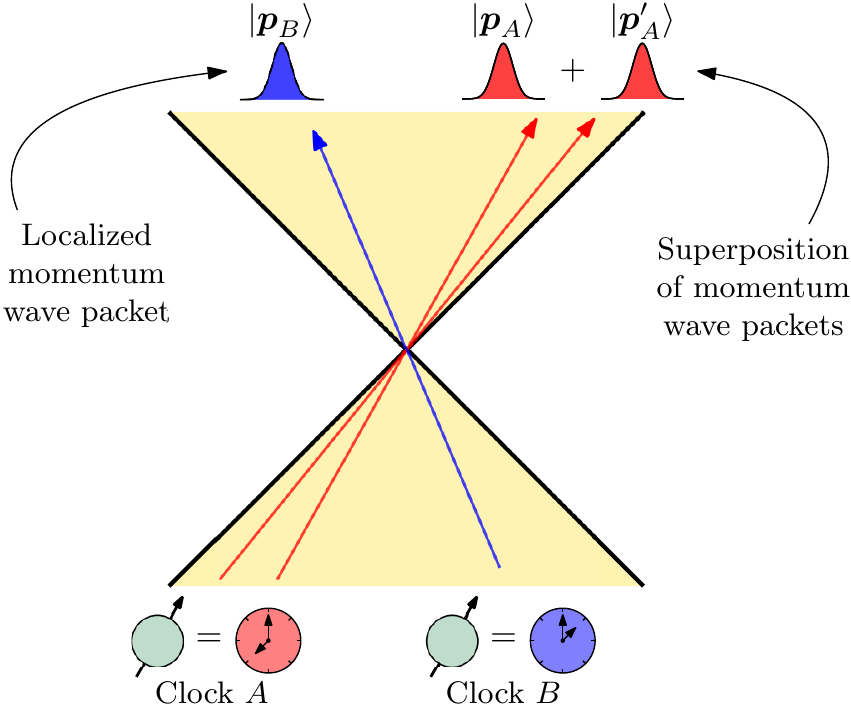} 
\caption{The world lines of clock $A$ (red) and clock $B$ (blue) moving through Minkowski space. Clock $A$ moves in a superposition of momentum wave packets giving rise to the quantum time dilation effect relative to clock $B$, which moves in a localized velocity wave packet.
  }\label{Fig1}
\end{figure} 
 
\begin{figure}
\includegraphics[width= .47 \textwidth]{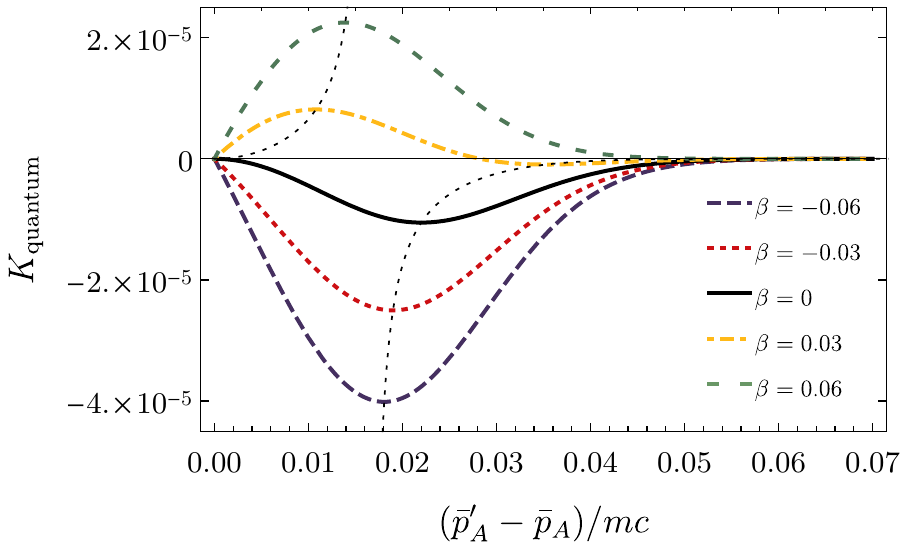}
\caption{ The dependence of the magnitude $K_{\rm quantum}$ of the quantum time dilation effect in one-dimension as a function of the difference $(\bar{p}_A' - \bar{p}_A)/mc$ of the average momentum of each wave packet comprising the superposition in Eq.~\eqref{supState} for different values of their sums $\beta \ce (\bar{p}_A + \bar{p}_A')/mc$ and \mbox{$\Delta/mc =0.01$}. The dotted black line traces the optimal difference in average velocity of the wave packets.
}\label{Fig2}
\end{figure}

\begin{figure*}[t]
\includegraphics[width= .8\textwidth]{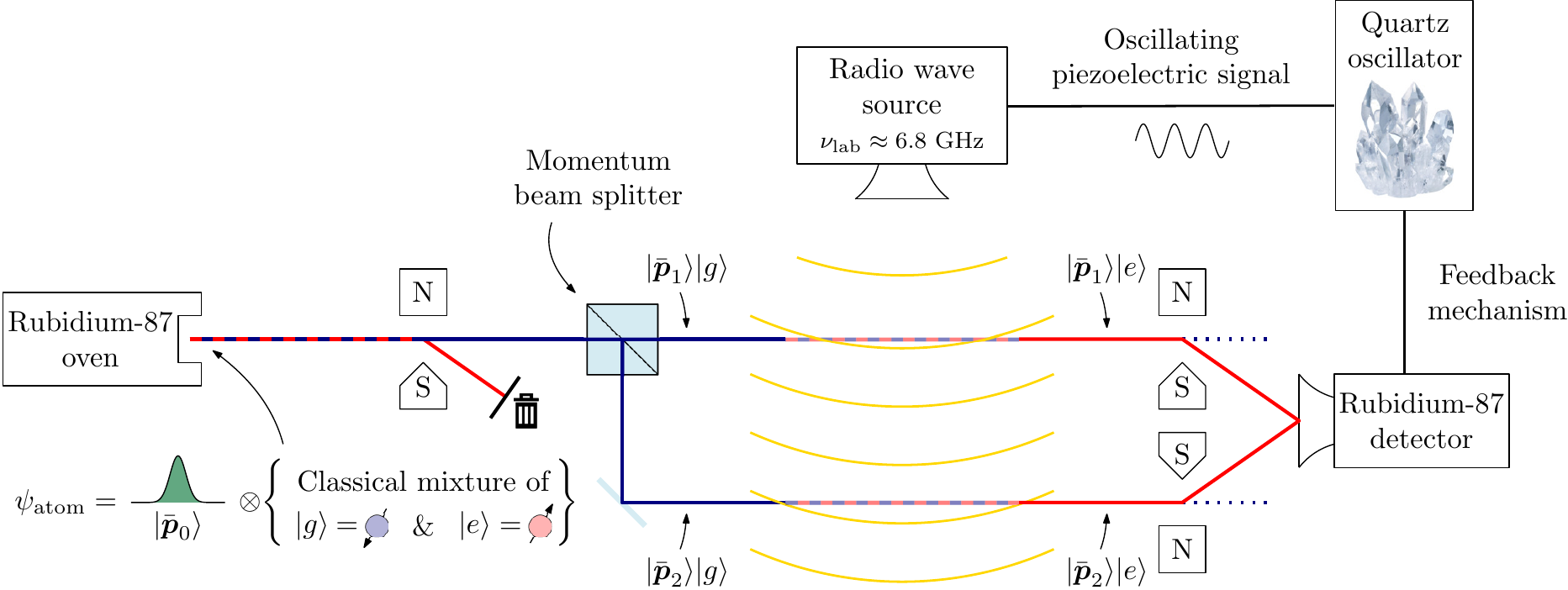} 
\caption{An experimental setup designed to observe the quantum time dilation effect based upon a modified atomic clock. Rubidium-87 atoms emerge from an oven, and then pass through a set of magnets to select only unexcited states of the atom $\ket{g}$ (blue line). These atoms then pass through a momentum beam splitter, emerging in a superposition of momentum wave packets with average momenta $\bar{\boldsymbol{p}}_A$ and $\bar{\boldsymbol{p}}_A'$; see Eq.~\eqref{supState}. This atomic superposition then passes through a radio source tuned at a frequency $\nu_{\rm lab}$ causing  the atoms to transition to their exited state $\ket{e}$ (red line), and thus be detected. The frequency $\nu_{\rm lab}$ is tuned  to maximize the number of atoms the detector registers. This will occur when  $\nu_{\rm lab}$ is in resonance with the frequency $\nu_{g \to e}$ associated  with the atomic transition $\ket{g} \to \ket{e}$. This resonance frequency is Doppler shifted so that its optimal value is $\nu_{\rm lab} = \gamma_{\rm eff}^{-1} \nu_{g \to e}$, which is seen to depend on $K_{\rm quantum}$, thus serving as a signature of the quantum time dilation effect. To examine the absorption rate in such an experiment a model of the light-matter interaction that takes into account the coherent delocalization of the atoms' momenta is required~\cite{stritzelbergerCoherentDelocalizationLightMatter,Smith:2020}. A related  experiment was recently described in Ref.~\cite{paigeClassicalNonclassicalTime2020}.} \label{Fig3}
\vspace{-10pt}
\end{figure*}

Having  derived the probability that $A$ reads the proper time $\tau_A$ conditioned on $B$ reading the time $\tau_B$ in Eq.~\eqref{probability}, we can answer the questions posed earlier. Suppose that $\ket{\psi_A^{\rm ext}}$ and $\ket{\psi_B^{\rm ext}}$ describe wave packets localized around different relativistic velocities  $\boldsymbol{v}_A = \bar{\boldsymbol{p}}_A/\gamma_A m$ and  $\boldsymbol{v}_B = \bar{\boldsymbol{p}}_B/ \gamma_B m$, where $\gamma_i \ce  \sqrt{1+\bar{\boldsymbol{p}}_i^2/m^2 c^2}$, and the associated internal clocks are ideal.\footnote{More precisely, the clock Hilbert space is taken to be $\mathcal{H}_C \simeq L^2(\mathbb{R})$, the clock Hamiltonian to be the momentum operator $\hat{H}_C=\hat{P}_C$, and the proper time observable $T_C$ is associated with the `position operator' $\hat{T}_C$ canonically conjugate to the momentum operator $[\hat{T}_C,\hat{H}_C] = i$. It follows that the clock states $\ket{\tau}$ correspond to eigenstates of $\hat{T}_C$. We emphasize, however, that the framework introduced can handle a wide variety of more realistic clocks.} For such clocks, the average time dilation of $A$ relative to $B$ given by the probability distribution in Eq.~\eqref{probability} agrees  with the prediction of special relativity, $\braket{T_A} = \frac{\gamma_B}{\gamma_A} \tau_B$. However, suppose that instead the center-of-mass of $A$ is prepared in a superposition of relativistic velocities  $\boldsymbol{v}_A = \bar{\boldsymbol{p}}_A/ \gamma_A m$ and  $\boldsymbol{v}_A' = \bar{\boldsymbol{p}}_A'/ \gamma_A' m$, so that
\begin{align}
\ket{\psi_A^{\rm ext}} = \frac{1}{\sqrt{N}} \left( \cos \theta \ket{\bar{\boldsymbol{p}}_A} + e^{i \phi}\sin \theta  \ket{\bar{\boldsymbol{p}}_A '} \right),
\label{supState}
\end{align}
where $N \ce \braket{\psi_A^{\rm ext}|\psi_A^{\rm ext}}$, and $\ket{\bar{\boldsymbol{p}}_A}$ and $\ket{\bar{\boldsymbol{p}}_A'}$ denote Gaussian wave packets in momentum space centred  around $\boldsymbol{\bar{p}}_A$ and $\boldsymbol{\bar{p}}_A'$ with a width $\Delta$; this situation is depicted in Fig.~\ref{Fig1}. In this case, the average time dilation of clock $A$ relative to clock $B$  to leading order in $H_C,  \tfrac{\boldsymbol{p}^2}{2m}\ll mc^2$ is 
\begin{align}
\braket{T_A} = \gamma_{\rm eff}^{-1} \tau_B = \left(1 - K_{\rm classical} - K_{\rm quantum} \right) \tau_B,
\nn 
\end{align}
where $\gamma_{\rm eff}^{-1} \ce 1 - K_{\rm classical} - K_{\rm quantum} $ is an effective relativistic factor that is a combination of classical and quantum contributions:
\begin{align}
K_{\rm classical} 
&\ce    \frac{ \bar{\boldsymbol{p}}_A^2 \cos^2 \theta +  \bar{\boldsymbol{p}}_A'^2 \sin^2 \theta - \bar{\boldsymbol{p}}_B^2}{2m^2 c^2 },
\nn \\ 
K_{\rm quantum} &\ce \frac{  \sin 2 \theta \cos \phi }{8  m^2 c^2N}   e^{- \frac{\left(  \bar{\boldsymbol{p}}_A' - \bar{\boldsymbol{p}}_A \right) ^2}{4\Delta^2}}  \nn \\
&\quad \times\!
\Big[   2\left(  \bar{\boldsymbol{p}}_A'^2 - \bar{\boldsymbol{p}}_A^2 \right) \cos 2 \theta - \left(\bar{\boldsymbol{p}}_A' - \bar{\boldsymbol{p}}_A \right)^2    \Big]. 
\nn 
\end{align}
The term $\tau_B K_{\rm classical} $ is the classical contribution to the average time dilation of clock $A$ relative to $B$ that would be observed if $A$ was prepared in a classical  mixture of momentum wave packets with average momenta  $\bar{\boldsymbol{p}}_A$  with probability $\cos^2 \theta$ and $\bar{\boldsymbol{p}}_A'$ with probability $\sin^2 \theta$. The term $\tau_B K_{\rm quantum} $ quantifies a nonclassical contribution to the time dilation of $A$ relative to $B$ that stems from the quantum nature of the momentum superposition in Eq.~\eqref{supState}. We refer to nonzero $\tau_B K_{\rm quantum} $ as \emph{quantum time dilation} because this contribution vanishes for classical mixtures of momentum wave packets. The magnitude of the quantum time dilation effect is plotted in Fig.~\ref{Fig2}. One might imagine extending this analysis to clocks moving in superpositions of accelerations and making connections with recent investigations into quantum aspects of the equivalence principle~\cite{zychQuantumFormulationEinstein2018,anastopoulosEquivalencePrincipleQuantum2018,giacominiQuantumMechanicsCovariance2019}.

The non-vanishing magnitude of $K_{\rm quantum}$ indicates the existence of a quantum time dilation effect as a result of clock $A$ moving in a superposition of velocities. An experimental setup is described in  Fig.~\ref{Fig3}  that could in principle observe  a signature of nonzero $K_{\rm quantum}$. This setup makes use of a momentum beam splitter~\mbox{\cite{bermanAtomInterferometry1997,cladeLargeMomentumBeamsplitter2009}} to prepare the momentum superposition in Eq.~\eqref{supState} and relies on maximizing the transition probability of an atom by tuning incident radiation, as in a standard atomic clock. State of the art experiments can observe classical special relativistic time dilation using optical atomic clocks moving at speeds as low as 10\,m/s~\cite{reinhardtTestRelativisticTime2007, chouOpticalClocksRelativity2010}. Moreover, velocity superpositions of Rubidium-87 wave packets moving at these speeds have been prepared using a momentum beam splitter realized by exploiting coherent momentum exchange between atoms and light~\cite{bermanAtomInterferometry1997,cladeLargeMomentumBeamsplitter2009}. Clocks moving in a superposition of velocities on the order of 10\,m/s results in the magnitude of the quantum time dilation effect to be $ \tau_B K_{\rm quantum} \approx  \tau_B  10^{-15}$. Given the current $10^{-14}$\,s resolution of atomic clocks~\cite{camparoRubidiumAtomicClock2007}, provided the coherence time of the velocity superposition is approximately 10\,s, which has been accomplished in the experiments of Kasevich \emph{et al.}~\cite{kovachyQuantumSuperpositionHalfmetre2015}, then it is expected that a signature of the quantum time dilation effect can be observed in a setup similar to that depicted in Fig.~\ref{Fig3}.

In conclusion, we can now answer the questions raised at the beginning in the affirmative. \emph{There exists a novel contribution to the time dilation observed by a clock moving in a quantum superposition of two relativistic  velocities when compared to an ensemble of clocks moving in a classical mixture of the same two velocities.} Given the current resolution of atomic clocks, it is argued that it is likely  possible to observe this quantum time dilation effect with current atomic physics techniques, thus offering a new test of relativistic quantum theory in the regime where strong coherence exists across relativistic momentum eigenstates.

\emph{Acknowledgments}: I thank Mehdi Ahmadi for our enjoyable  collaboration. This work was supported by the Natural Sciences and Engineering Research Council of Canada and the Dartmouth College Society of~Fellows. 

\onecolumngrid
\vspace{-.1in}
\twocolumngrid

\bibliography{GRFEssay2020}

\end{document}